\documentclass{aastex631}

\shorttitle{Growth and Decay of Sunspots and Pores}
\shortauthors{Tlatov }
\graphicspath{{./}{figures/}}

\begin{document}

\title{The Growth and Decay of Individual Sunspots and Pores}

\author[0000-0002-6286-3544]{Andrei G. Tlatov}
\affiliation{Kislovodsk Mountain Astronomical Station of the Pulkovo observatory, Kislovodsk, Gagarina str. 100, 357700, Russia}

\begin{abstract}
An analysis of the photometric growth and decay rates of sunspots and pores was carried out.
According to the \textit{Solar Dynamics Observatory/Helioseismic and Magnetic Imager} (SDO/HMI) data for the period 05.2010-03-2025, $\approx 3.5\cdot 10^{\rm 5}$  sunspots and pores were detected and their evolution tracked.   The growth and decay rates of sunspots depend non-monotonically on the area.  For small-area sunspots $S \lesssim 50$ $\mu$sh, a rapid increase in velocity is observed with increasing the area for the growth stage $dS^{\rm gr}_{\rm sp}\approx 0.2 \cdot S^{\rm 1.35}$ and for the decay stage $dS^{\rm dc}_{\rm sp}\approx 0.28 \cdot S^{\rm 1.26} $  $\mu$sh/day.  For sunspots $S\gtrsim 50$ $\mu$sh, the growth rate depends weakly on the area: $dS^{\rm gr}_{\rm sp}\approx 11.9 \cdot S^{\rm 0.32}$ $\mu$sh/day.  For the decay stage of sunspots with an area of $S\approx 50\,-\,150$ $\mu$sh, the decay rate also depends weakly on the area and can be approximated as $dS^{\rm dc}_{\rm sp}\approx 11 \cdot S^{\rm 0.15} $. For sunspot areas $S\gtrsim 150$ $\mu$sh, the decay rate accelerates with increasing area: $dS^{\rm dc}_{\rm sp}\approx 0.14 \cdot S^{\rm 0.96} $ $\mu$sh/day. For solar pores, the growth and decay rates of solar pores are linearly related to the area. The growth and decay rates for the spots of the leading and trailing polarities are determined. In the range of spot areas $S\approx 100\,-\,200$ $\mu$sh, the growth and decay rates of sunspots of trailing polarity are higher than those of sunspots of leading polarity.

\end{abstract}


\section{Introduction} \label{sec:intro}
The life cycle of sunspots can be divided into a growth stage and a decay stage. The growth rate of sunspots is usually much higher than the decay rate. \cite{Bumba} found that the sunspot growth rate is $A\approx t^{\rm 4}$, while the decay rate for sunspots with an area of $A=100$ millionths of a solar 
hemisphere ($\mu$sh, 1 $\mu$sh=$3.32\ \ Mm^{\rm 2}$) $day^{\rm -1}$ is proportional to $A\approx A_{\rm 0} exp(-t/t_{\rm 0})$.  \cite{Howard} found for sunspot groups that the growth rate averages $\approx 502 \%$/day, while the decay rate is $\approx 45\%$/day, i.e. the growth and decay rates depend on the area of the sunspot groups.  At the same time \cite{Dalla} estimate the growth rate as an area-independent value of $\approx 30\,–\,40$ $\mu$sh/day. \cite{Javaraiah} found that growth rates are approximately 70\% higher than decay rates.

Sunspot decay plays an important role in the transfer of the solar surface magnetic flux. One of the features of decay is that the area of sunspot groups $A$ decreases linearly with time. This pattern is related to the Gnevyshev\,-\,Waldmeier (G\,-\,W) rule   \citep{Gnevyshev, Waldmeier} and has been confirmed in other studies (see, for example, \cite{Bumba, Gokhale}). In \citep{Meyer} it was found that the average decay rate for 60-day spots is constant for most of their life $dA/dt\approx -3\cdot 10^{\rm 12}$ $cm^{\rm 2}/s \approx  0.26\  Mm^{\rm 2}$/day. \cite{Bumba} obtained a linear area decay rate of -4.2 $\mu$sh/day for recurrent sunspots. \cite{Pillet} found a mean decay rate of -12.1 $\mu$sh/day for sunspot groups. \cite{Li} 
found that the decay rates of $\alpha$ configuration sunspots range from -7.75 to -23.81 $\mu$sh/day. Other authors also noted a linear dependence of the decay rate \citep{Gafeira, Muraközy20}.

\cite{Hathaway} found that the decay rate of sunspot groups has a linear dependence on area for sunspot groups $S<1000$ $\mu$sh, but decreases for groups $S>1000$ $\mu$sh. Also they pointed out the problems of studying the speed of 
determination associated with the non uniformity of data series of the RGO 
($1874\,-\, 1976$) and USAF (after 1976) data series.

Several authors associate sunspot decay with a parabolic process \citep{Moreno, Pillet, Petrovay, Litvinenko}, where the temporal change in the sunspot area is proportional to the square root of the area. Some researchers believe that the decay rate is different for leading and trailing sunspots \citep{Howard, Muraközy20}. \cite{Howard} noted that small sunspot groups grow on average faster than they decay, while large sunspot groups decay on average faster than they grow. In \citep{Muraközy20}, based on the study of 206 different active regions, the decay rate of the trailing parts was found to be higher than that of the leading parts in the sunspot group. \cite{Tlatov23} showed that the duration of the growth and decay stages of sunspots depends on the polarity of the magnetic field. For the same area, the decay stage of sunspots of the leading polarity has the longest duration. For solar pores, the polarity of the magnetic field does not matter.

The nature of the decay, linear or proportional to the sunspot perimeter, determines two main decay models: turbulent diffusion and turbulent erosion. The turbulent diffusion model assumes that sunspot decay is independent of sunspot size and that the loss of magnetic flux occurs everywhere in the sunspot \citep{Meyer, Pillet}. In the turbulent erosion model, sunspot decay is considered to occur near sunspot boundaries and is correlated with the perimeter \citep{Simon, Petrovay, Litvinenko}.

In this study, we investigate the growth and decay rates of sunspots and pores using large statistical data.  We will also consider the differences in the growth and decay rates of sunspots depending on their polarity.

\section{Data and Processing Methods}
Sunspots last from a few hours to several months \citep{Vitinskij}. To measure the growth and decay rates of small area sunspots, several observations per day must be used. For the analysis, we used a database of individual sunspot characteristics obtained from SDO/HMI observations of the hmi.Ic\_45s and hmi.M\_45s series, performed at the same time. We took 5 images for each day at times close to 00:00, 05:00, 10:00, 15:00, and 20:00 UT (\url{http://jsoc.stanford.edu/ajax/exportdata.html}). 

The main data for processing were images in “white” light (hmi.Ic\_45s), in which sunspots and sunspot umbra were highlighted. To identify sunspots and pores, we used the sunspot boundary detection procedure \citep{Tlatov14, Tlatov19, TT24}.   In this analysis, we used the characteristics of sunspots and pores in the period 01.05.2010\,-\,29.03.2025. More than 108 thousand sunspots and more than 250 thousand pores were identified. Using these data, we tracked the evolution of spots by constructing sequences of chains.

To form chains, we used the procedure of tracking sunspots selected in neighboring time images. To search for spots, we tracked spots that had the same magnetic polarity, were at a minimum heliographic distance, and differed little in area. To do this, we search for the spots in neighboring images and have a minimum of the parameter: $d_{\rm ij}=\sqrt{(\phi_{\rm i}-\phi_{\rm j})^{\rm 2}+(\theta_{\rm i}-\theta_{\rm j})^{\rm 2}}+(\Delta S)^{\rm 2}$, where $\theta,\phi$ are the latitude and longitude of the spots $i,j$. We corrected the longitude of the spots on the subsequent image for differential rotation $\phi_{\rm j}=\phi_{\rm j}'-(13.198+\Delta t/24 + \Delta \omega)$,  where $\Delta t$  is the time between observations in hours, the term $\Delta \omega$ takes into account the differential rotation in latitude $\Delta \omega=0.44-3 \Delta t sin^{\rm 2}\theta$. The functional of the area was calculated as: $\Delta S =3 abs(S_{\rm i}-S_{\rm j})/(0.5(S_{\rm i}+S_{\rm j})+20+2 \Delta t)$, here $S_{\rm i}$ and $S_{\rm j}$ are the area of spots, expressed in millionths of a hemisphere ($\mu$sh).  The spots were considered identical if their coordinates were less than $|\theta_{\rm i}-\theta_{\rm j}|<2+\Delta \theta/2$  and $|\phi_{\rm i}-\phi_{\rm j}|<2+\Delta \phi/2$, where $\Delta \theta$, $\Delta \phi$ are the extent of the spots in latitude and longitude, and the minimum of the parameter $d_{\rm ij}$ was found for the pair ${ij}$. If spot j could not be found for spot $i$ in the adjacent image, the sequence chain was interrupted. In total, more than 80 thousand chains of individual spots were identified.

\begin{figure}[ht!]
\centerline{\includegraphics[width=0.8\textwidth,clip=]{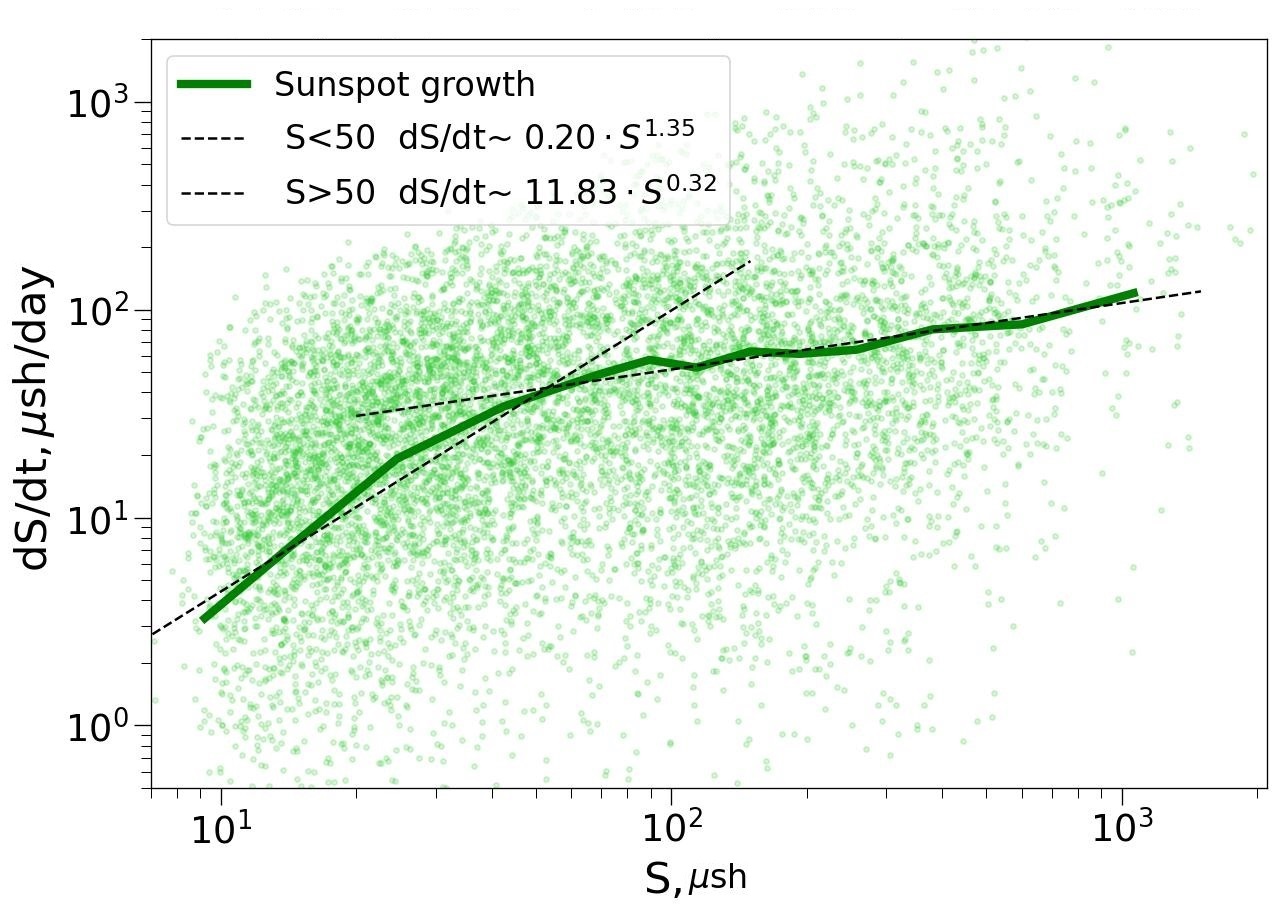}
                 } 
\caption{Distribution of sunspot growth rates depending on the area. The green line shows the average values. The dotted lines are the approximation lines of the different sections.}
\label{fig:fig1}
\end{figure}

\begin{figure}
\centerline{\includegraphics[width=0.8\textwidth,clip=]{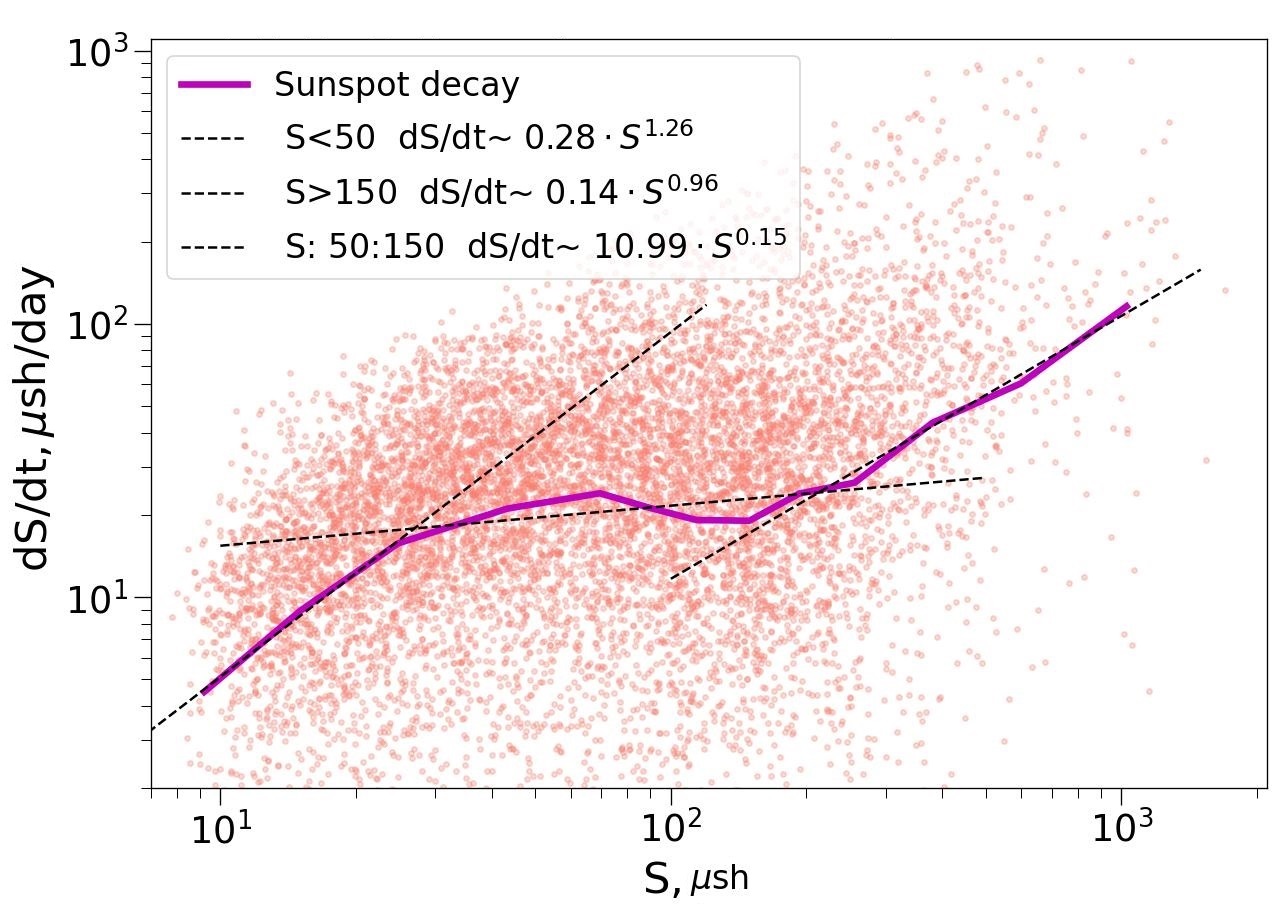}
                 } 
\caption{Distribution of sunspot decay rates. The lines represent average values. The dotted lines are approximation lines.}
\label{fig:fig2}
\end{figure}

\section{Rate of Change of Sunspots and Pores}  
\subsection{Rate of Growth and Decay of Sunspots and Pores}
To avoid artifacts on the solar limb to determine the rate of area change in our study, the location of sunspots was limited to $r<0.6R_{\odot}$.  In this section, we present the rate of change of sunspots in images separated by $\approx 5$ hours. To determine the growth and decay stages in the sequence chains, we determined the time $t_{\rm mx}$ when the largest area was reached $S_{\rm mx}$.  Then we define the growth stage as $t<t_{\rm mx}$ and the decline stage as $t>t_{\rm mx}$. To determine the growth rate, we calculated the changes in the area for each spot $dS^{\rm gr}/dt=(S_{\rm l+1}-S_{\rm l})/\Delta t$ for the growth stage  $t_{\rm mx}$ and $dS^{\rm dc}/dt=(S_{\rm l}-S_{\rm l+1})/\Delta t$. To form the average values, we took into account all changes in the area, even in the growth stage $dS^{\rm gr}/dt <0$, or in the decline stage $dS^{\rm dc}/dt >0$.

Figure \ref{fig:fig1} shows the scatterplot, average values for area intervals, and approximations for the sunspot growth phase as a function of the sunspot area. Here, by sunspots we mean spots with a penumbra. The dots represent approximately 10\% of all spots, selected at random. The dependence of the growth rate is nonuniform. For sunspot areas lower  $S<50$ $\mu$sh, the growth rate is observed depending on the area $dS^{\rm gr}_{\rm sp}/dt \approx 0.2\cdot S^{\rm 1.35}$  $\mu$sh/day. For sunspots with an area $S>50$ $\mu$sh, the growth rate depends weakly on the area: $dS^{\rm gr}_{\rm sp}/dt \approx 11.9\cdot S^{\rm 0.32}$ $\mu$sh/day. So for sunspots with an area $S\approx 100$ $\mu$sh, the growth rate is $dS/dt\approx 50$ $\mu$sh/day.

The decay rate of sunspots is shown in Figure \ref{fig:fig2}. The dependence of the decay rate of sunspots on the area can be divided into 3 sections. For small spots $S<50$ $\mu$sh, the dependence can be described by the expression:  $dS^{\rm dc}_{\rm sp}/dt \approx 0.28\cdot S^{\rm 1.26}$   for $S: 50\,-\,150$ $\mu$sh:  $dS^{\rm dc}_{\rm sp}/dt \approx 11\cdot S^{\rm 0.15}$ $\mu$sh/day. For large sunspots $S\gtrsim 150$ $\mu$sh  the dependence can be expressed as: $dS^{\rm dc}_{\rm sp}/dt \approx 0.14\cdot S^{\rm 0.96}$ $\mu$sh/day.  For a sunspot with an area of $S\approx 100$ $\mu$sh, the decay rate will be: $dS^{\rm dc}_{\rm sp}/dt \approx 20$ $\mu$sh/day.

\begin{figure}
\centerline{\includegraphics[width=0.8\textwidth,clip=]{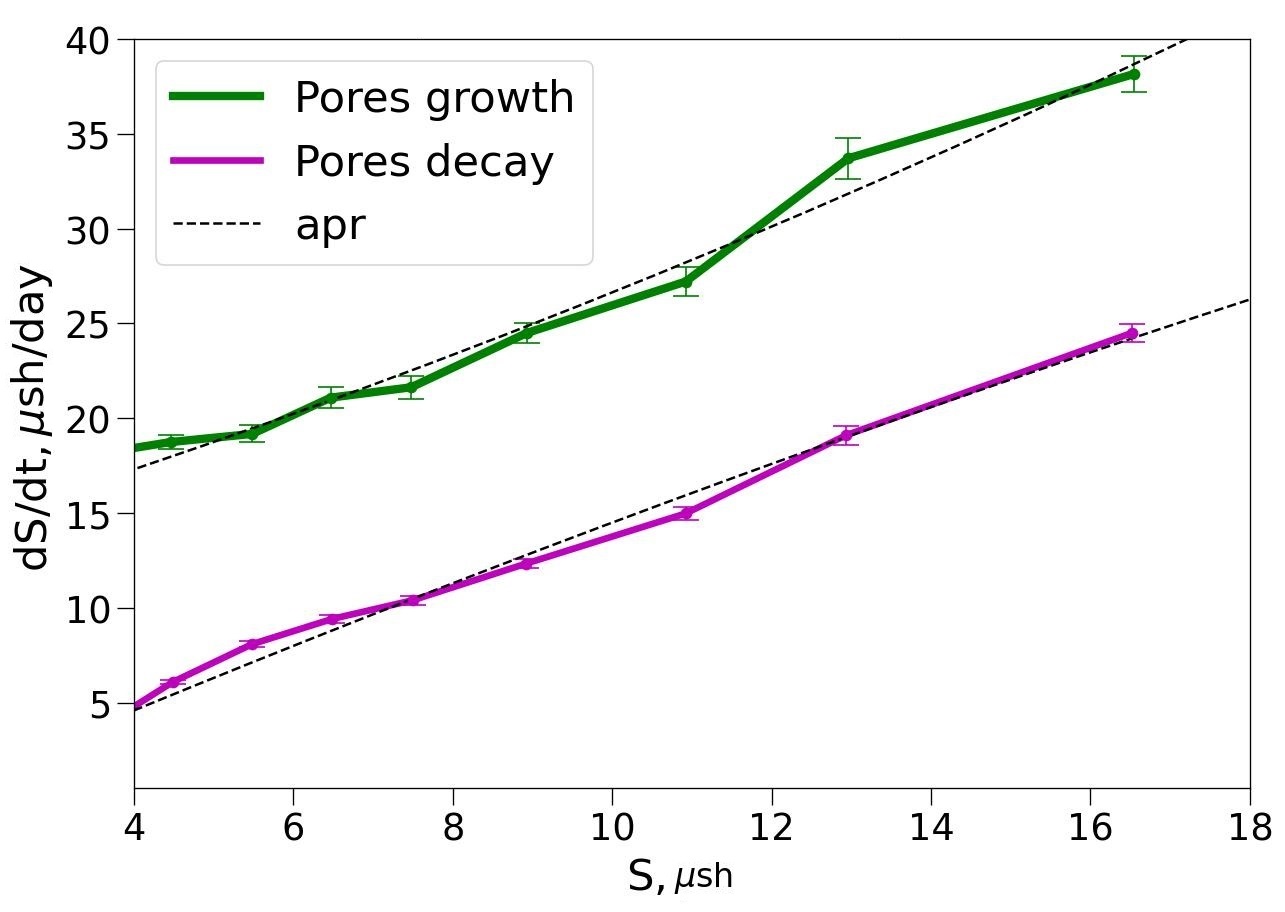} } 
\caption{Growth and decay rates for solar pores. Dotted lines represent approximations. Bars represent confidence intervals.
}
\label{fig:fig3}
\end{figure}

The rates of area change for solar pores differ from those for sunspots. Unlike sunspots, pores do not have a penumbra. As a rule, these are spots with an area of less than $S\approx 10\,-\,15$ $\mu$sh. The dependences of the growth and decay rates for pores are shown in Figure \ref{fig:fig3}.  In addition to the average value in the selected area intervals, the confidence interval $\sigma/\sqrt{n+1}$ is also presented, where $n$ is the number of samples in the interval. For the growth rate, the dependence can be represented as  $dS^{\rm gr}_{\rm pr}/dt \approx 19+1.2\cdot S$  $\mu$sh/day; for decay $dS^{\rm dc}_{\rm pr}/dt \approx-2.5+1.8\cdot S$  $\mu$sh/day.

\begin{figure}
\centerline{\includegraphics[width=0.8\textwidth,clip=]{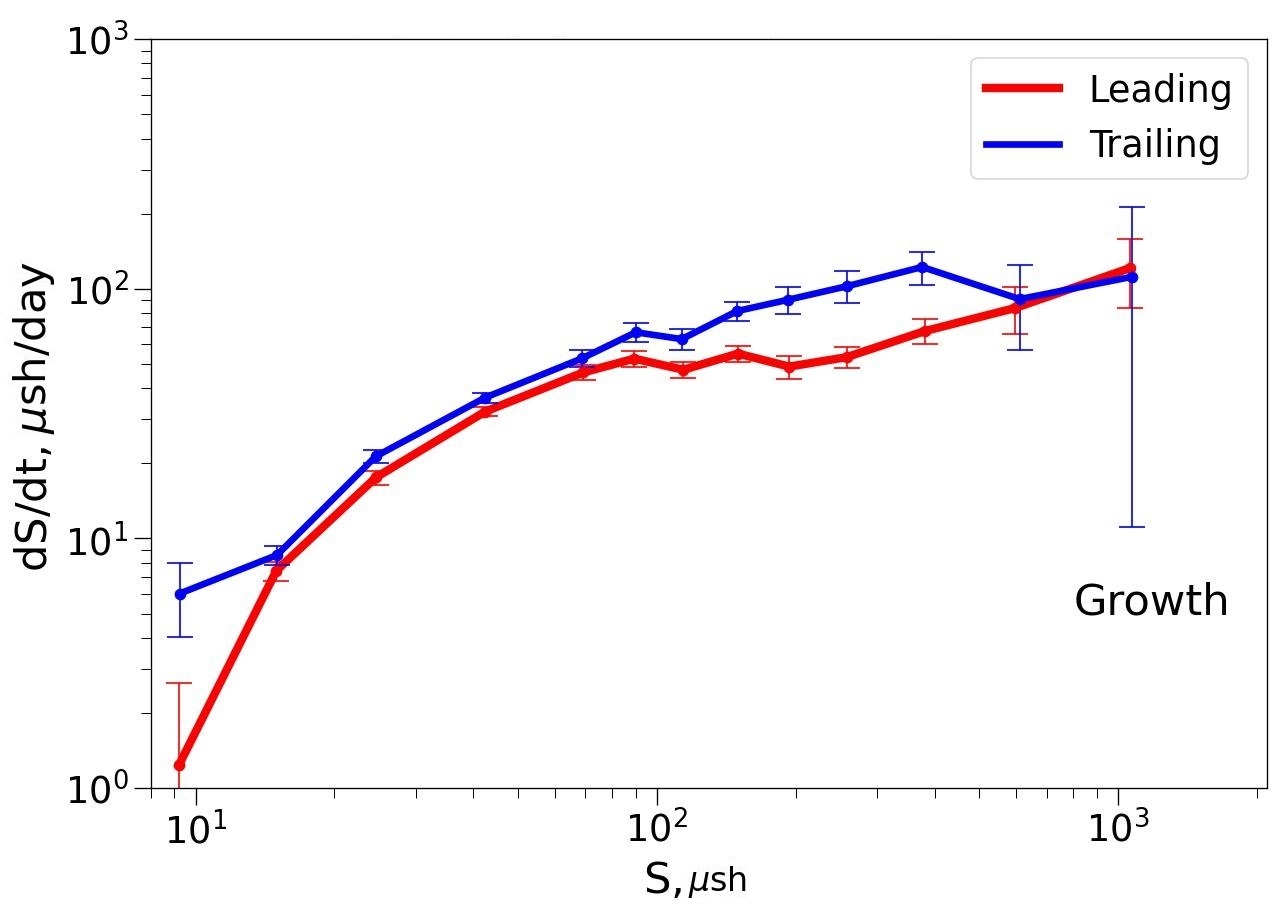} } 
\caption{Growth rate of sunspot and pore areas for spots of leading and trailing polarity. 
}
\label{fig:fig4}
\end{figure}

\begin{figure}
\centerline{\includegraphics[width=0.8\textwidth,clip=]{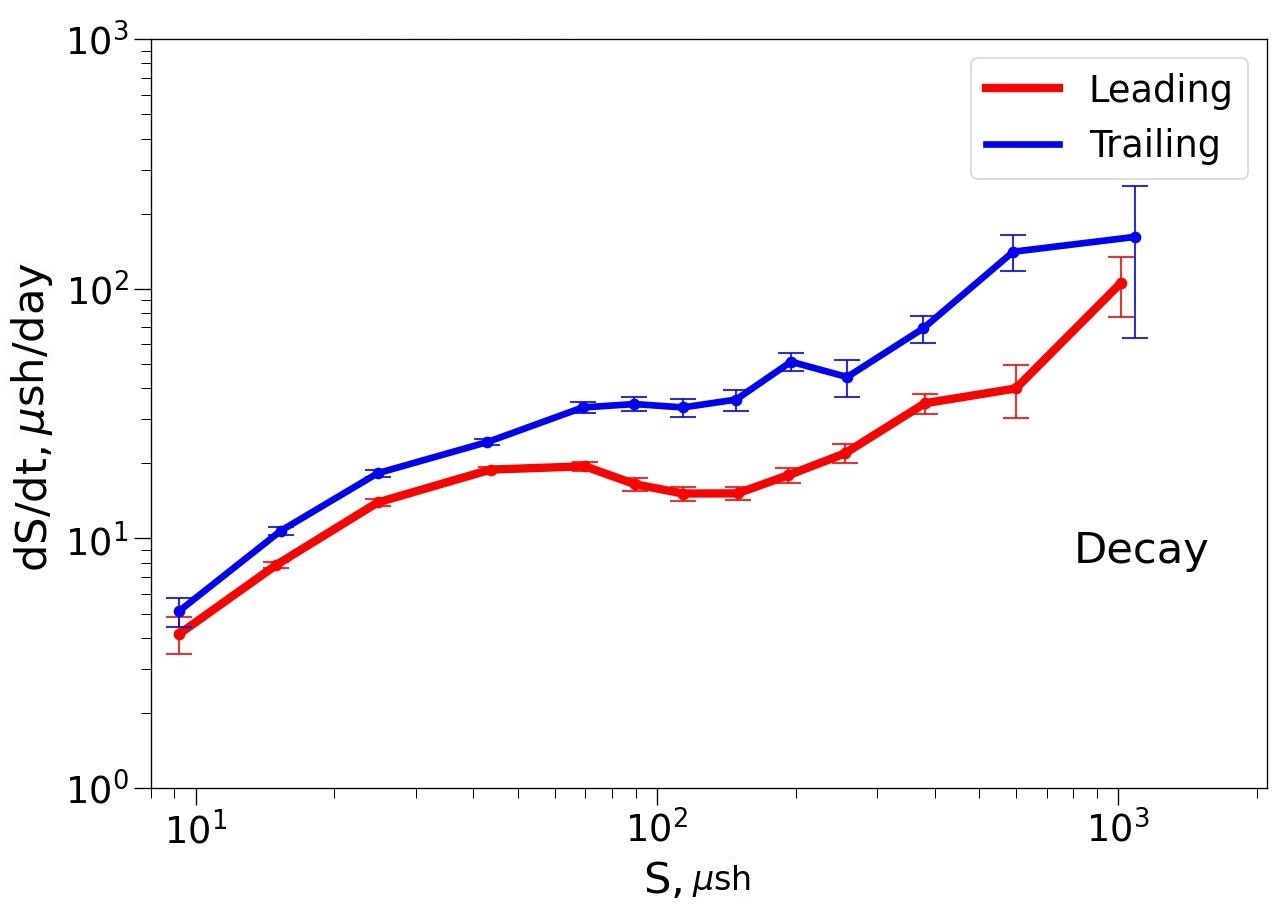} } 
\caption{Same as in Figure \ref{fig:fig4} but for the decay stage.}
\label{fig:fig5}
\end{figure}

\subsection{Rate of Change  of Sunspots of Leading and Trailing Polarity}
We superimposed the boundaries of sunspots and pores on the magnetic field images obtained with the SDO/HMI instrument at the same time as the continuum observations. This allowed us to divide the spots into spots of leading and trailing polarity in accordance with Hale's rule. In total, 122,585 pores and 74,383 sunspots of leading polarity and 127,789 pores and 34,356 sunspots of trailing polarity were identified. The mean area of sunspots of the leading polarity was $S^{\rm ld}_{\rm sp} =143$ $\mu$sh, in the trailing polarity $S^{\rm tr}_{\rm sp} =104$ $\mu$sh.
Figure \ref{fig:fig4} shows the dependences of the growth rate of sunspot and pore areas for spots of leading and trailing polarity. The growth rate of area for sunspots of trailing polarity is higher than that for sunspots of leading polarity. This is especially noticeable for sunspots with an area of more than $S>100$ $\mu$sh.   For sunspots with an area of $S :50\,-\,200$ $\mu$sh, the growth rate of sunspots of leading polarity is $dS^{\rm gr}_{\rm ld}/dt \approx30$ $\mu$sh/day and  $dS^{\rm gr}_{\rm tr}/dt \approx 80$ $\mu$sh/day for sunspots of trailing polarity.
The decay rate of sunspots of leading and trailing polarity is shown in Figure \ref{fig:fig5}. Trailing polarity sunspots decay faster than leading ones. For sunspots with an area of $S:50\,-\,200$ $\mu$sh, the decay rate of leading polarity from the area becomes almost constant:  $dS^{\rm dc}_{\rm ld}/dt \approx 10$ $\mu$sh/day and for sunspots of trailing polarity $dS^{\rm gr}_{\rm tr}/dt \approx 30$ $\mu$sh/day. For solar pores, no difference in the growth and decay rates could be detected in spots of different polarity.

\section{Discussion}
We performed an analysis of the photometric growth and decay rates of sunspots and pores using white-light observations during the 24th and early 25th solar activity cycles. We extracted sunspot and pore boundaries using 5 images per day. The photosphere-penumbra boundary was superimposed on magnetic-field maps made with the same telescope at the same time. This allowed us to track the evolution of more than $3\cdot 10^{\rm 5}$ sunspots and pores over time, taking into account magnetic polarity. We found that the growth and decay rates depend non-monotonically on the area. For pores and sunspots of a small area $S < 50$ $\mu$sh, the growth and decay rates are approximately proportional to the first power of the area. For sunspots $S > 50$ $\mu$sh the dependence of the growth and decay rates changes significantly. For the growth rate, the dependence on the area is proportional to the power of $\approx 0.3$ (Figure \ref{fig:fig1}). The decay rate of sunspots practically does not change with the area (Figure \ref{fig:fig2}).

\begin{figure}
\centerline{\includegraphics[width=0.8\textwidth,clip=]{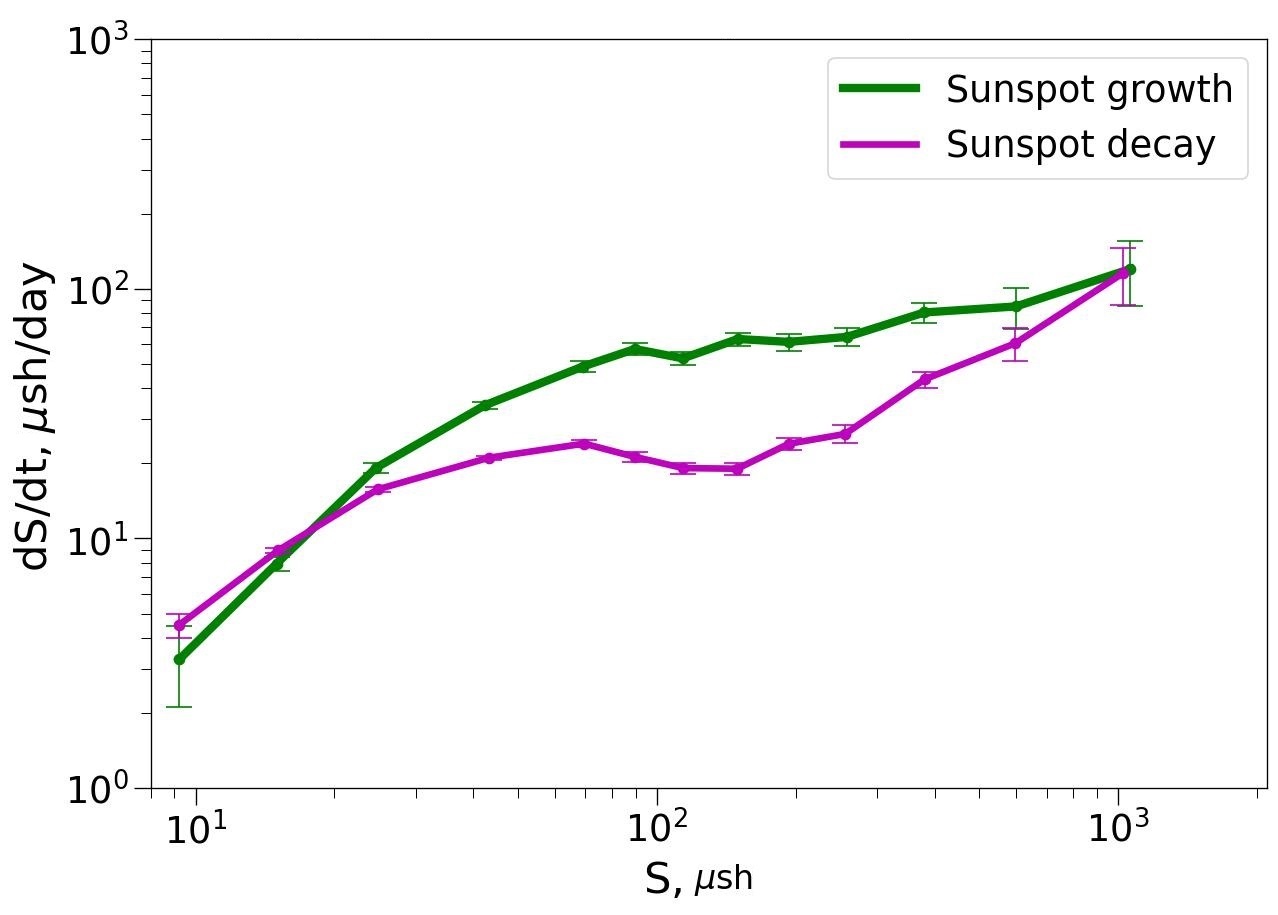} } 
\caption{Dependences of growth and decay rates on sunspot area. Confidence interval  are presented.}
\label{fig:fig6}
\end{figure}

\begin{figure}
\centerline{\includegraphics[width=0.8\textwidth,clip=]{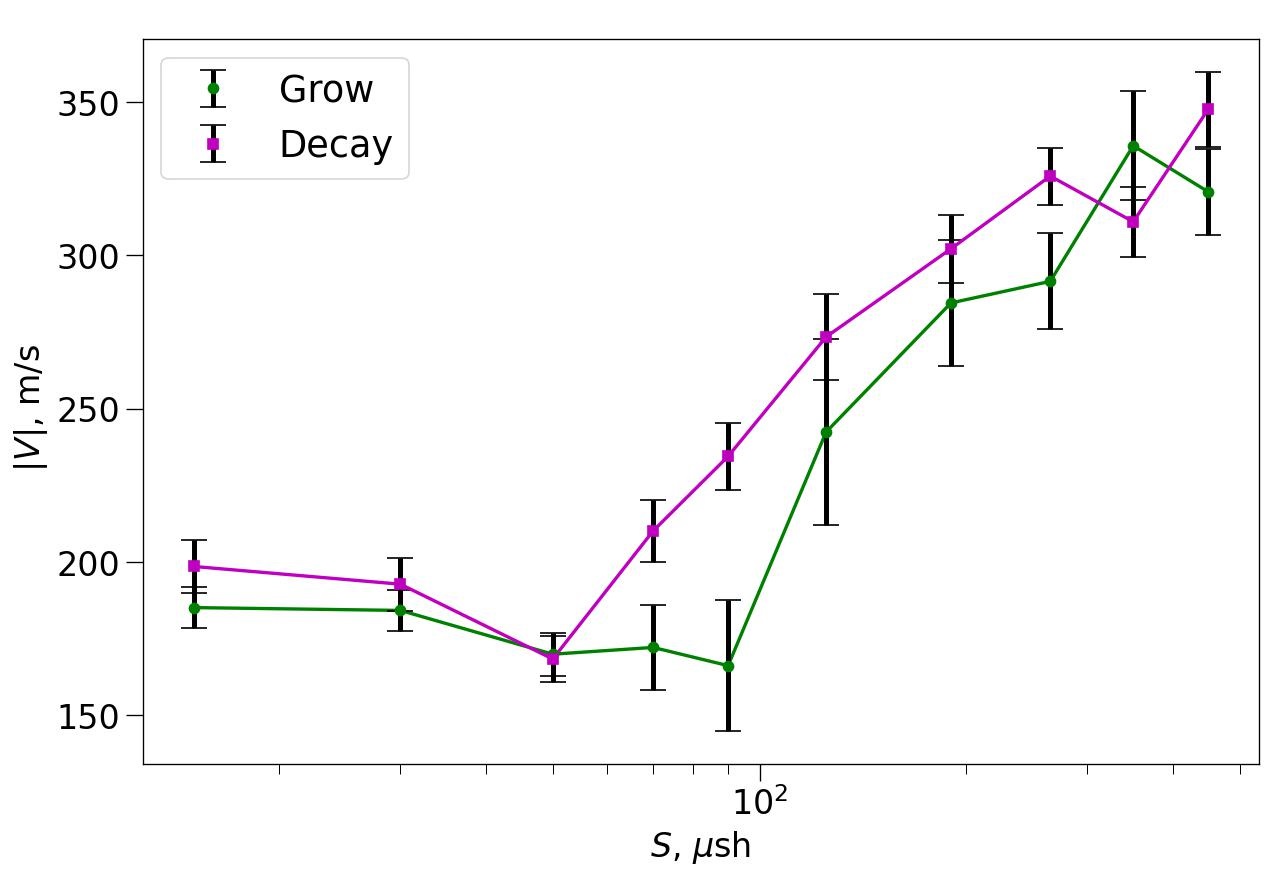} } 
\caption{Absolute velocities of matter in sunspots, observed along the line of sight for spots in the central zone of the solar disk $r<0.4R_{\odot}$. Dependences on the area of spots are presented for the growth and decay stages.}
\label{fig:fig7}
\end{figure}

In Figure \ref{fig:fig6}, we combine growth and decay rates. The composition of the growth and decay rates is similar to that of hysteresis. For areas larger than $S\gtrsim 10^{\rm 3}$ $\mu$sh, the decay rate is comparable to the growth rate. This may limit the area of sunspots to a few thousand $\mu$sh.

The values of the growth rate found $dS^{\rm gr}_{\rm ld}/dt \approx 30$ $\mu$ sh / day and $dS^{\rm gr}_{\rm tr}/dt \approx 80$ $\mu$ sh / day are close to the results obtained by \citep{Dalla}. The decay rate $dS^{\rm dc}_{\rm ld}/dt \approx 10$ $\mu$sh/day $\approx 3$ M$m^{2}$/day for sunspots and  $dS^{\rm dc}_{\rm tr}/dt \approx 30$ $\mu$sh/day $\approx  10$ M$m^{2}$/day. The rate values for sunspots are close to the Gnevyshev-Waldmeier rule $\approx 11$ $\mu$sh/day and the data on the decay of long-lived sunspots \citep{Meyer}. Note that in our analysis, we used all sunspots, not just long-lived ones.

The graph of growth and decay rates in Figure \ref{fig:fig6} resembles hysteresis. The hysteresis effect in sunspots occurs when modeling the formation of the penumbra \citep{Rempel}. However, it is possible that the dependence in Figure \ref{fig:fig6} is not related to the “memory” effect characteristic of the description of hysteresis, but to the velocity of matter in sunspots. We superimposed sunspot boundaries on the SDO/HMI hmi.V.45s LOS velocity maps at 5:00 UT. We preliminarily subtract differential rotation and orbital velocities from the velocity maps. For this analysis, we used data at a short distance from the solar disk center $r<0.4R_{\odot}$. In this way, we mainly extracted vertical velocities in sunspots. Figure \ref{fig:fig7} shows the unsigned velocities $|V|$ for sunspots at the growth and decay stages. As in Figure \ref{fig:fig6}, for areas $S: 50\,–\, 500$ $\mu$sh, a difference is visible for growth and decay. Higher vertical velocities are observed at the decay stage. In a study of the lifetime of sunspots (Tlatov, 2023), different dependencies were also found for small-area spots ($S<50\,-\,100$ $\mu$sh) and larger sunspots. It was suggested that for small sunspots and pores, photometric disappearance is associated with the heating process and can be described by Newton's formula. In this case, in addition to a decrease in the magnetic field, the temperature in the spot equalizes, and the rate of equalization is proportional to the temperature or area itself. For spots with a larger area, the velocities of matter in the penumbra of sunspots become important. The decay and growth of sunspots with an area of $S>50$ $\mu$sh is possibly associated with the velocity of matter \citep{Tlatov23}.

The Hale polarity of the magnetic field is important for sunspots with an area of $S\gtrsim 50$ $\mu$sh. The decay rate of sunspots of the trailing polarity is $\approx 2\,-\,3$ times higher than the decay rate of sunspots of the leading polarity (Figure \ref{fig:fig5}).

The non-monotonic nature of the sunspot decay rate obtained by us can help in understanding the contradictory results of the data on the sunspot decay rate. The decay of sunspots with an area of $S: 50\,-\,200$ $\mu$sh generally satisfies the G-W rule, i.e. it is constant or weakly dependent on the area. For pores and sunspots of small area and for sunspots of large area the decay rate is approximately proportional to the first power of the area (Figures \ref{fig:fig2}, \ref{fig:fig3})     This is probably due to the peculiarities of the formation of the sunspot penumbra for this range of areas (\cite{Tlatov23}, Fig. 9) and the peculiarities of the formation of the magnetic field (\cite{Tlatov23}, Fig. 7,8). 

The lifetime of sunspots is determined by the duration of the decay stage. We have shown that the decay rate varies non-monotonically, there are area ranges ($S<50$ $\mu$sh and $S>200$ $\mu$sh) when the decay occurs quickly, approximately proportional to the first power of the area. But in the area range $S\approx 50\,-\,200$ $\mu$sh, the decay of sunspots occurs slowly, practically not changing with the area (Figure \ref{fig:fig2}). This area range corresponds to the maximum of the area distribution \citep{Solanki, Tlatov19}. Perhaps this is why some researchers \citep{Gnevyshev, Bumba} recorded a linear dependence of the decay rate of sunspots on time. Outside this range, the dependence of the decay rate on the area changes its character. In this case, other researchers recorded power dependences on time \citep{Meyer}. It follows from this that the description of the decay process of sunspots within the framework of one model is not applicable.

From the decay rate analysis in Figure \ref{fig:fig2} it follows that for a sunspot with an area of $S\approx 100$ $\mu$sh the characteristic decay rate will be  $dS^{\rm dc}_{\rm sp}/dt \approx 20$ $\mu$sh/day. This is approximately twice as much as for the sunspot groups of the same area according to the G-W rule. This means that the lifetime of sunspot groups is approximately 2 times longer than the lifetime of individual spots of the same area. This result is consistent with the analysis \citep{Tlatov23}. Probably, the lifetime of sunspot groups is also determined by the appearance of other sunspots in the group, and not by the lifetime of the largest sunspot.

\begin{acknowledgments}
We acknowledge the financial support of the Ministry of Science and Higher Education of the Russian Federation, grant number 075-03-2025-420/4.

\end{acknowledgments}

\bibliography{spot_decay.bib}{}
\bibliographystyle{aasjournal}

\end{document}